# COMPONENT BASED BUSINESS SOLUTIONS UNDER ARCHITECTURE

T.A. Barrett  
Origin  
Groningensingel 1  
6835 EA Arnhem  
The Netherlands  
Tim.Barrett@nl.origin-it.com

H.A. Proper  
Origin  
Hoogoorddreef 5  
1100 AN Amsterdam  
The Netherlands  
E.Proper@acm.org

## 1. Introduction

Many of today's applications have an, almost tangible, monolithic nature. They are built as 'islands', purporting to be self contained, offering little or nothing in the way of integration with other applications. In the past, being large and self-contained may have eliminated the need to interact with other solutions to some extent. However, in the business environments of today the interaction with other applications becomes paramount. As a result of this, many ad-hoc point-to-point integration solutions have been built between different applications. This has already led to an 'application spaghetti' at many of our customer sites. Many of today's applications are poorly structured, which makes their responsiveness to business change sluggish. The application spaghetti with its plethora of point-to-point interfaces further inhibits the responsiveness to change.

The use of component based technologies carries the promise of providing more flexible IT solutions. The use of these technologies is expected to further enable re-use, lead to an improved integration of off-the-shelf components, as well as the integration of legacy applications. Adapting the use of component-based technology, however, does introduce some new challenges. As argued in e.g. [1], information systems in the future are expected to take the form of a 'swarm' of business objects. These business objects correspond to the "things" around which a business is organised. They may be grouped into (or simply treated as) components; leading to 'swarms' of components.

When the information systems of the future indeed turn into swarms of components, a pressing practical issue seems to be how to maintain control over these swarms. Some questions that may be raised are:

- What business functionality is currently supported?
- How can we meet new business requirements?
- How can we enable new business?
- How can we ensure a high level of flexibility?
- How can we integrate existing legacy systems? Should we?
- How can we integrate off-the-shelf applications or components? Should we?
- How can we interface with other enterprises?

The rigidity of the monolithic applications of old, at least gave businesses some feeling of command over their systems portfolio. Components provide more flexibility, but may be harder to control.

In this paper we present a two pronged approach to tackle these issues. Firstly, we outline an architectural approach to the development of component-based business solutions. Secondly, we propose a reference architecture to help in actually realising such solutions. The architectural approach to system development aims to provide a way to provide better control of a components environment.

## 2. Business Solutions under Architecture

Gaining control over the development directions of component based solutions requires a thorough understanding of both the business and the technology context of these solutions. This need comes even more to the fore when we consider the conditions under which most organisations currently have to operate. These conditions exhibit a high degree of dynamism. A dynamism that forces them to become more flexible: "evolve or die". Reduced protectionism, deregulation of international trade, de-monopolisation of markets, privatisation of state owned companies, increased global competition, cross-border merges, the emergence of trade blocks, the shift from a single economic power to a multi-polar economic world, new currencies, are all aspects of this increasingly dynamic business environment [2, 3]. In other words, the domains in which organisations operate continuously evolve, and the speed at which they evolve is still increasing; evolution is a constant.

Tapscott [2] proposes an architectural approach as a way to improve the flexibility of organisational structures and IT in particular. The Gartner Group has also published numerous reports on the need for architectural approaches. In [4, 5], and more recently in [6], the Gartner Group identified the need for a shift in IT paradigms. This shift is needed to meet the need for more flexibility of the IT architecture. A shift is needed from a *technology-centred* to a more *application-centred* approach. The general message of these architecture developments is that IT should *empower* a business with the means to go out and seek new challenges.

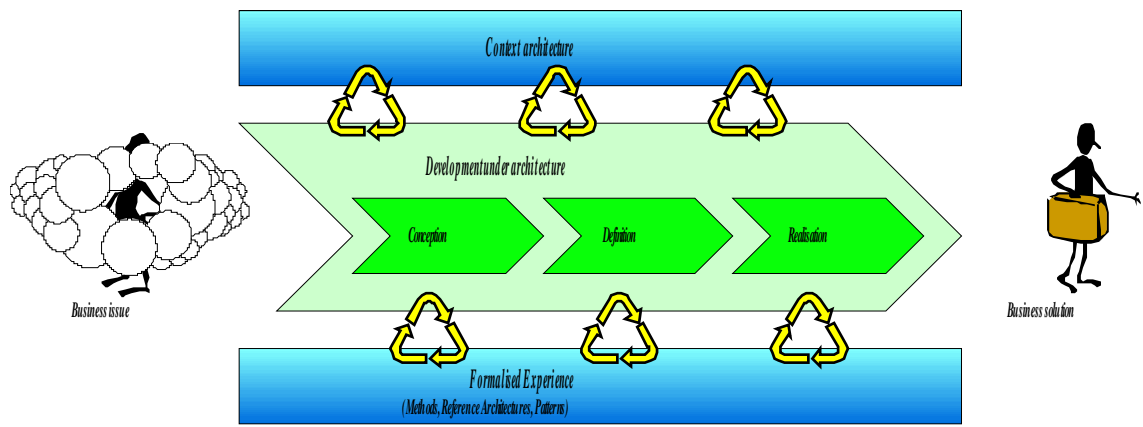

*Figure 1: Development under architecture*

Developing component based information systems in such rapidly evolving context is like shooting at a moving target [7-9]. Gaining a good understanding of the business context, and the direction in which it is likely to evolve, will make it easier to align business and IT. In [10, 11], it is also argued that business-IT alignment is a continuous process, where business is a driving force and IT an enabling force.

Our view of the development of component based business solutions under architecture is illustrated in Figure 1. Development under architecture consists of three major phases. The whole process starts with a business issue. In the conception phase, this business issue is properly identified. This consists of a problem definition, an analysis of the stakeholders, and an analysis of the current situation. During the definition phase, the desired solution (architecture) is defined that fits within the provided context architecture. The aim of the realisation phase is migrate from the current situation to the desired situation. This latter phase in particular may lead to a multitude of smaller sub-projects that each try to realise part of the overall effort.

The entire development effort is supported by two important classes of input:

- *Context architecture.* The solution that will be devised needs to fit within a certain architectural context; i.e. *under architecture*. This context may be the current business and technology context of the solutions under development, but can equally be a long-term vision. Compare this to the way the architectural design of a new office building fits within the zoning plan of the area in which the building will be erected.

- *Experience.* Experience from the past will help us in building new solutions. A lot of experience will be in the heads of those who do the actual development. However, an increasing amount of experience can be formalised and captured as:
  1. *Methods.* A development methodology can be seen as a formalisation of experience. When applying a method in practice, new experience will be gained, which may lead to refinements of the pre-existing method.
  2. *Reference architectures & patterns.* It is well known that for specific classes of problems generic solutions can be constructed that can be tailored to more specific situations [12]. This may range from the re-use of generic patterns and components, to the application of complete reference architectures. In the next section we will discuss one such reference architecture.

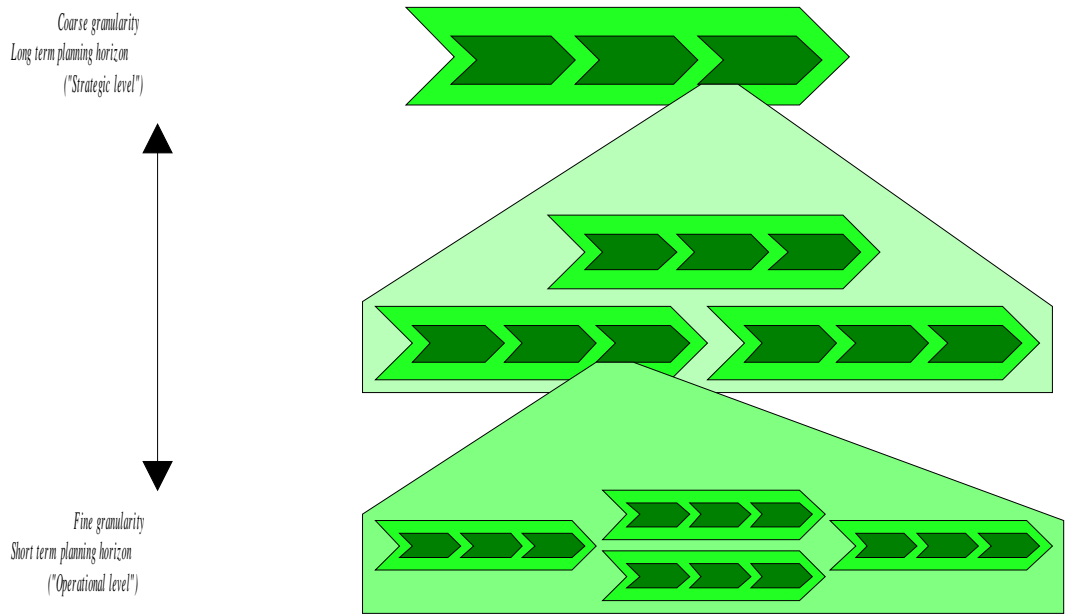

*Figure 2: Granularities and planning horizon*

The situation depicted in Figure 1, can be repeated on multiple levels of granularity and time horizons. This is illustrated in figure 2. At the top level, we may have a project which is concerned with a high level architectural view of the business solutions to be developed. This project may give rise to some more concrete projects that are aimed at filling in some aspects of the high level architecture. At the bottom of the chain, we can find the projects that actually build the component-based solutions. Each of the projects at a finer level of granularity are responsible for filling in part of the architecture as defined by the coarser grained architecture.

Now how does this improve an organisation's ability to control large numbers of components? The current and desired architectures at each of the different levels of granularities and planning horizons provide the insights needed in making decisions about the structuring of the components and other infrastructure issues. This may concern such issues as:

- **The structuring of components**. For example, when an insurance company knows that it will regularly update its insurance products, it is important to set-up the component architecture such that new products may be introduced without causing changes to the structure or content of existing components. Another example would be a bank, that has decided to offer its products through a variety of distribution channels (bank offices, call-centre, internet). In this latter case it may be worthwhile to identify core components that are generic with respect to the distribution channel used.

- **The integration of legacy applications**. If a legacy application exists that provides the functionality needed by some component, it may be worthwhile to use the existing legacy application (using a wrapper) for the time being.
- **Make-or-buy decisions**. Buying off-the-shelf components is probably cost effective in the short term. For the longer term this depends on the flexibility of the off-the-shelf component. If changes in the required functionality are expected, off-the-shelf components may not be able to meet these changes.

## 3. The Technical Reference Architecture

In this section we discuss a reference architecture that can be used as a base to build component based solutions. The architectural approach decribed above can be used as a method for driving the precise structuring of a particular solution conforming to this reference architecture. The Technical Reference Architecture (TRA) provides:

- A typology of the layers into which the internal components of a solution are divided, and the responsibilities of the components within these layers.
- A detailed definition of the way in which component based solutions interact with legacy applications.

The TRA is a reference architecture for **Component Based Transactional Solutions.** Other solution styles exist and reference architectures would need to be defined for these. The term *transactional* is used here as an umbrella term. Essentially we are saying that *enterprise strength* solutions involve the manipulation of large amounts of data by large numbers of concurrent users. This data is a vital business resource, and its integrity must be guaranteed. This means that concurrent access to data must be managed, security must be managed, and data cannot be allowed to become inconsistent. *Transactions* ensure that changes to data are treated as units of work, which either succeed as a whole or are backed out in their entirety, as if the changes had never taken place. A TP monitor provides these resource management functions, as well as providing a powerful, scalable run time environment.

Component based transactional solutions are solutions with the fundamental features of enterprise scale solutions. They are built using layers of components - many of these will be candidates for reuse. We are unwilling to use the term application, as applications tend to be large, and self contained, and can often only interact with other applications with difficulty. Instead, we prefer to use the term *logical component*. A logical component provides a solution to a particular business problem. We would expect to see a 'customer information' logical component for example. Logical components are themselves built up of layers of finer grained components, as described below. Logical components have well defined external interfaces that describe the services made available to other logical components.

A final feature of TRA is the fact that the layering scheme within each logical component provides vital abstraction behind which various degrees of implementation freedom can be exercised. This means that integration with legacy applications is hidden to a large extent, and the logical components have a large amount of freedom regarding the resource managers used (database management systems used, message queuing technology etc.), as well as the *presentation technology* used.

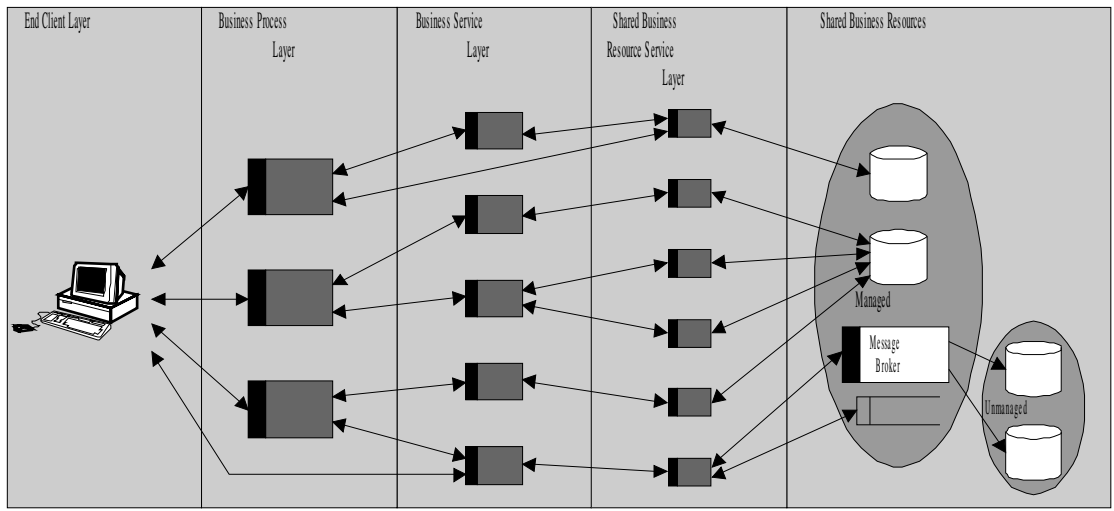

*Figure 3: Technical Reference Architecture overview*

Figure 3 provides a high level overview of the Technical Reference Architecture, which is explained in more detail below. The Technical Reference Architecture covers two major areas. These are *construction* and *integration*. Construction and Integration are technical implementation exercises. The identification and modelling of components is the responsibility of the higher-level business oriented architectures described in section 2.

### 3.1 Construction

The basis of TRA is *layering*. The layers have the following typology:

- End Client Layer; The components in this layer are primarily concerned with presentation services, and contain little or no business logic. These components are 'end clients'; they are also 'thin clients'.

- Business Process Layer; Services exist within this layer which support the definition, composition, assignment and execution of business processes. This is a highly dynamic layer that directly supports dynamic real life business processes. Layers below this are less dynamic, and more concerned with Business Resource (data) integrity.

- Business Service Layer; A business service component represents a specific, explicitly modelled, recognisable business entity, such as Customer, Product, Contract etc. Components belonging to this layer are responsible for driving or participating in recoverable 'ACID' *transactions* that update multiple shared business resources. They are expected to run in an environment that can handle distributed transactions. This environment can pass transaction contexts onto other components that are not necessarily running in the same process as the transaction originator. Transaction co-ordination is a vital feature, usually provided by TP monitors. Business services also provide *business rules*, explicitly provided via service interfaces, as well as *algorithms.* Business Service components delegate all data manipulation work to components in the Shared Business Resource Service Layer.

- Shared Business Resource Service Layer; The components within this layer provide basic read, update, create and delete services related to specific, explicitly modelled, recognisable business entities such as Customer, Product, Contract etc. This layer is sometimes referred to as a data abstraction layer, as it hides physical databases (or interaction with legacy applications) from the business service components. As a result of this abstraction, business service components interact with data and legacy applications purely in terms of the interfaces provided by components in the Shared Business Resource Service Layer. This gives the client business service components a great deal of platform and environment independence.

- Shared Business Resource Layer; This layer contains physical *data*, held in databases, files, message queues, etc. Note that resources fall into two categories:
  - **Managed Resources:** This refers to data held in databases, or being transmitted in message queues which is directly owned by the higher level components.
  - **Unmanaged Resources:** This refers to data held in existing 'legacy' applications. The higher level components can use these resources, but cannot control them directly in a transactional sense.

In both cases, the concept of a *resource manager* is of fundamental importance. A resource manager is able to commit or rollback its resources under the control of the component which is co-ordinating a transaction. Such a transaction may affect multiple resource managers. We are in fact describing a 2-phase commit protocol. Transaction co-ordination of this type cannot usually extend to include legacy applications. Full transaction co-ordination across logical components is fully addressed by the architecture however. This is depicted in figure 4, where we see a business service component within one logical component driving a transaction which includes the invocation of a shared business resource layer component in another logical component. Legacy applications can also be resource managed *indirectly*. Most message queuing products (certainly those from IBM and Microsoft) are themselves resource managers. This means that if a message queue is used to send a message to a message broker, then the *actual sending* of this message will only occur at the *commit point*, when all updates to local managed resources have succeeded.

## 3.2 Integration

Integration is a vital concern. In the case of logical components interfacing with other logical components, there is no need to engineer interfacing in as an 'afterthought'. Logical components are in fact made up of components offering services, some of which can be invoked by external clients. In cases involving legacy applications it is highly unlikely that the legacy applications will have been designed with a standardised form of interfacing. The ability to make their services available to external client applications has to be added on, as described below.

### 3.2.1 Message Brokers

A message broker allows a legacy application to project services expressed via interfaces. This means that a component wishing to use a service provided by a legacy application need only to invoke the desired service, as defined in the interface for that service held in the message broker. The 'new' component does not need to deal with the legacy application directly, instead it deals with the service interface projected by the message broker. New components are not 'polluted' with intimate knowledge of the back-end legacy applications.

The message broker constructs messages that can be understood by the target legacy applications. Enterprise strength message brokers are usually table driven, which means that the messages to be created and sent to legacy applications are defined in tables as opposed to being hard coded in programs. Fully-fledged message brokers are also *engines*, in the sense that they are capable of sending multiple messages, in parallel if possible, to multiple back-end applications in order to service a request. The responses received from the legacy applications are translated and sent back to the requesting component.

## 3.3 Adapters

Figure 4 depicts two logical components interacting with one another directly, using agreed upon, standardised protocols. These components however are both interacting with legacy applications via the interfaces projected by a message broker.

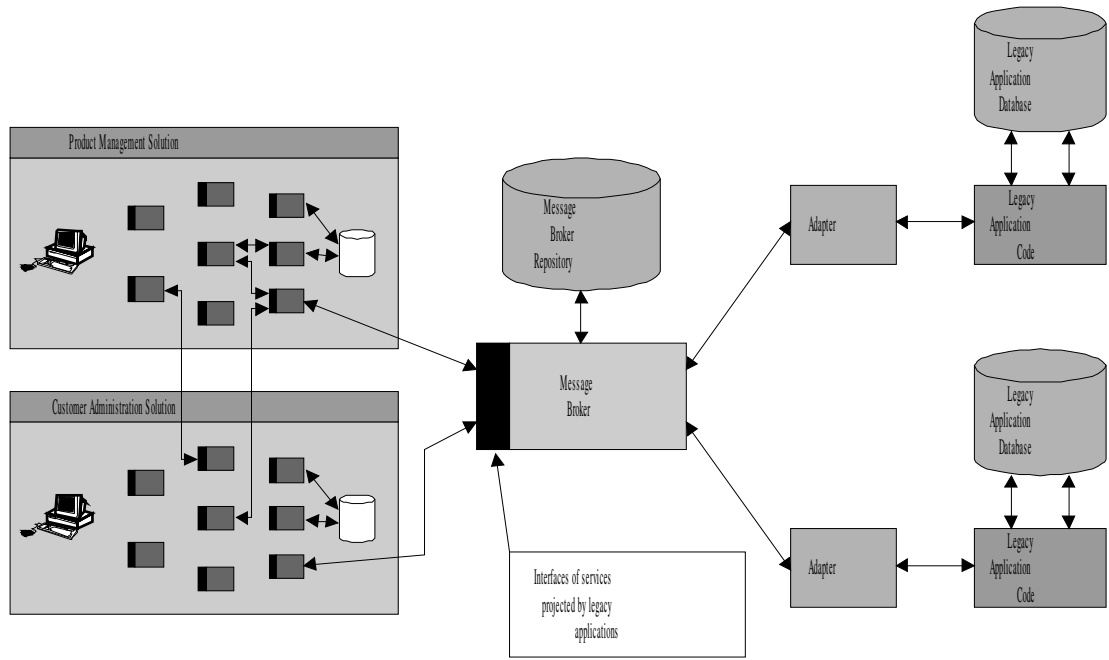

*Figure 4: Message broker enabled integration*

Figure 4 depicts three vital aspects of the technical reference architecture:

1. The existence of logical components
2. The internal structure of these logical components - layers of internal components
3. Native integration between the logical components, and the use of message brokers and adapters to achieve integration with legacy applications.

## 4. Candidate Implementation Platforms

The Technical Reference Architecture (TRA) is implementation decision free. This means that the architecture is valid and valuable for a wide range of implementation platforms. By far the most important and difficult technical decision that must be made is the choice of the 'right' distributed component middleware and the programming model that (always) goes with it. The terms 'middleware wars' and 'object wars' are known throughout the industry. Being *implementation free*, the TRA is neutral in such wars. It provides a common sense layering and integration strategy which *encapsulates change* and balances *flexibility* with the key enterprise needs of *scalability*, *performance* and *business resource integrity*.

However, a reference architecture alone is not enough. IT development organisations must present *both* a consistent architecture *and* proven in depth skills in particular implementation environments. By specialising in selected implementation environments full teams of skilled people can be provided for customer projects, covering the full architectural spectrum from business context architecture to development and implementation. Specialising in the following vendor/implementation platform combinations will cover a vast section of the enterprise computing market:

- **Microsoft -** focusing on the **COM** programming model (in the near future COM+) supported by COM/DCOM, Microsoft Transaction Server, MSMQ, IIS and Dynamic HTML.

- **IBM** - focusing on (1) a 'traditional' environment concentrating on **CICS** on mainframes, Unix, As400 and NT, and (2) advanced (**CORBA/JAVA**) development using **Component Broker Connector**. MQSeries would provide a message queuing resource manager.

- **BEA** - again, focusing on the 'traditional' **Tuxedo** TP monitor product, and in the future the Object Transaction Monitor **Iceberg**. BEAMessageQueue would provide a message queuing resource manager.

- **Iona** - focusing on **Orbix/OTM and OrbixCOMet.** Although IONA is small compared to IBM and Microsoft, it nevertheless commands a great deal of market respect. IONA is also of interest because of its healthily objective attitude towards COM and CORBA; IONA has licensed the Microsoft COM source code and is developing OrbixCOMet which provides CORBA/COM interoperability.

- **Muscato Engin** - is a message broker which provides all of the features described in this paper.

The following key statement is made within [13]:

*"By 2000, the enterprise computing market will begin to settle on three fundamental environments: the OTM-based Active Server* **[Microsoft Transaction Server ed.],** *the OTM based Java/CORBA environment and mainframe CICS; more than 90 percent of all business transactions will run in these three environments (0.7 probability)"*

## 5. Conclusion

In this paper we have outlined an architectural approach to the development of component-based business solutions, and argued the importance of reference architectures in this context. Such reference architectures range from business level to implementation level, and typically revolve around multiple layers of abstraction.

We have concentrated on the discussion of technical reference architecture for component-based solutions. From a practical stance, we recognise the need to specialise on a limited number of strong technologies from major vendors. This will allow us to implement this architecture at customer sites.